\newcommand{\CLASSINPUTtoptextmargin}{0.98cm}
\newcommand{\CLASSINPUTbottomtextmargin}{0.6cm}

\documentclass[journal]{IEEEtran}
\IEEEoverridecommandlockouts                              

\usepackage{graphicx}
\usepackage{multirow}
\usepackage[table,xcdraw,dvipsnames]{xcolor}
\usepackage{epstopdf}
\usepackage{amsfonts}
\usepackage{amsmath,amssymb}
\usepackage{color,soul}
\usepackage{color}
\usepackage{verbatim}
\usepackage{comment}
\usepackage{algorithm}
\usepackage[table,xcdraw]{xcolor}
\usepackage{varioref}
\usepackage{gensymb}
\usepackage{comment}
\usepackage{xfrac}
\usepackage{soul}
\usepackage{color}
\usepackage{xcolor}
\usepackage{cite}
\usepackage{ctable}
\usepackage{MnSymbol}
\usepackage{dashrule}
\usepackage{relsize}
\usepackage{pifont}
\usepackage{bm}

\usepackage{ctable}

\usepackage{caption}
\usepackage{subcaption}
\usepackage{cleveref}
\definecolor{Orange1}{RGB}{250,235,215}
\newcommand{\norm}[1]{\left\lVert#1\right\rVert}

\begin{document}
	
	\title{Polyhedral Predictive Regions For Power System Applications}
	
	\author{Faranak~Golestaneh, 
		Pierre~Pinson,~\IEEEmembership{Senior Member,~IEEE,}
		 and~Hoay~Beng~Gooi,~\IEEEmembership{Senior Member,~IEEE}
	}
	\vspace{-0.8em}
	\maketitle
	
	\vspace{-0.3em}
	\begin{abstract}
Despite substantial improvement in the development of forecasting approaches, conditional and dynamic uncertainty estimates ought to be accommodated in decision-making in power system operation and market, in order to yield either cost-optimal decisions in expectation, or decision with probabilistic guarantees. The representation of uncertainty serves as an interface between forecasting and decision-making problems, with different approaches handling various objects and their parameterization as input. Following substantial developments based on scenario-based stochastic methods, robust and chance-constrained optimization approaches have gained increasing attention. These often rely on polyhedra as a representation of the convex envelope of uncertainty. In the work, we aim to bridge the gap between the probabilistic forecasting literature and such optimization approaches by generating forecasts in the form of polyhedra with probabilistic guarantees. For that, we see polyhedra as parameterized objects under alternative definitions (under $L_1$ and $L_\infty$ norms), the parameters of which may be modelled and predicted. We additionally discuss assessing the predictive skill of such multivariate probabilistic forecasts. An application and related empirical investigation results allow us to verify probabilistic calibration and predictive skills of our polyhedra.
	\end{abstract}


	%
	\IEEEpeerreviewmaketitle

	\vspace{-0.6em}
	\section{Introduction}

\PARstart{T}{he variability and} limited predictability of renewable power generation have introduced new challenges into power systems.
 With a large-scale uncertain generation, in order to reduce the gap between fail-safe and economical solutions of operational problems, advancement in two areas is essential. First, the development of highly scalable optimization techniques capable of accommodating considerable degree of uncertainty is required. Second, it is of utmost importance to develop adequate and high-quality representations of the uncertainties involved to be used as input to the aforementioned optimization techniques~\cite{bertsimas2013adaptive, bessa2017towards}.


Practitioners mostly use so-called deterministic or point forecasts as input to decision making today. These comprise single-valued prediction for the future realization of a variable of interest, disregard the actual range of potential outcomes. However, the solution to an optimization problem in a deterministic setup may be highly sensitive to small perturbations of uncertain quantities.  Hence, ignoring uncertainty can result in suboptimal or infeasible solutions in practice~\cite{ben2000robust}. 



Uncertainty forecasts can be represented in various forms such as scenario, probabilistic and ramp forecasts~\cite{morales2013integrating}. Since wind and PV power both show high cross-correlation in time and space, more recently, forecasting spatial/temporal scenarios has been of interest. For example, temporal uncertainty forecast is a key requirement for multi-period operational problems such as unit-commitment
and state of charge of energy storage~\cite{haessig2015energy}. 
Stochastic programming as one of the most common  optimization techniques in power systems applications uses scenarios as inputs to  find  optimal solutions in uncertain environments~\cite{lowery2015reserves,zheng2015stochastic}. However, stochastic programming holds a number of pitfalls in a practical context including heavy computational burden and the need for hard-to-obtain probability distributions~\cite{wiesemann2014distributionally}.


The issues with  stochastic programming motivates to move towards more  recent  approaches to optimization under uncertainty, namely robust, chance-constrained and interval optimization. Recently, these optimization techniques  have been deployed in power systems applications~\cite{doostizadeh2016energy,wei2016distributionally,hu2016robust }. For these classes of decision-making problems,  
the required uncertainty representation takes the form of prediction regions rather than scenarios. Robust optimization is a computationally viable methodology providing solutions deterministically immune to any realization of  uncertainty within a defined uncertainty set (another term for  prediction regions). Interval optimization  derives optimistic and pessimistic solutions based on boundaries of prediction regions. In chance-constrained optimization, the uncertainty sets give a probability guarantee for the coverage of observations from the stochastic process considered.

  Prediction regions in univariate case, e.g. modeling uncertainty of a single wind farm in a particular time, can be adequately addressed by prediction intervals~\cite{morales2013integrating}. However, when modeling temporal/spatial or multivariable correlations  is of interest, prediction regions take the form of multivariate ellipsoids, boxes and polyhedra. We  refer to uncertainty sets as prediction regions to emphasize on the fact that they are predictions in nature.


Although the multivariate prediction regions have been used in several optimization applications, the literature has been almost silent on how  to efficiently generate and evaluate them. The parameters of multivariate prediction regions are simply chosen based on assumptions or by trial-and-error without verification of those assumptions in practical applications. Uncertainty sets are constructed based on a Gaussian assumption in~\cite{hu2016robust} for nodal load and in~\cite{venzke2017convex,wei2016distributionally} for wind power. The inadequacy of a Gaussian assumption in describing uncertainty of wind and PV power is discussed in e.g.~\cite{golestaneh2018ellipsoidal}. A parameter named uncertainty budget is used to control the size and  conservativeness of wind power uncertainty sets in the form of ellipsoids in~\cite{li2015modeling} and in the form of polyhedra in~\cite{shao2017security}.  As a different approach, in~\cite{soares2017active} convex hull of spatial/temporal scenarios is defined as a prediction region of wind/PV. In~\cite{bessa2015marginal}, temporal scenarios are used as  input to produce multivariate prediction intervals (MPIs) to characterize the dependency of wind power forecast errors over a  time horizon.

Robust optimization tends to produce conservative solutions. The conservativeness of a robust solution is directly linked to the size of uncertainty sets
~\cite{zheng2015stochastic}. However, controlling size of uncertainty sets is not a trivial task to be determined arbitrarily. As any other type of prediction, uncertainty  prediction should provide a certain level of required performance. 
 Multivariate prediction regions are assessed based on their calibration and sharpness. Calibration is linked to conservativeness and it shows how close  the empirical coverage rate of a prediction region is  to its nominal one.  In contrast to~\cite{venzke2017convex,wei2016distributionally,hu2016robust,li2015modeling,shao2017security, soares2017active}, we emphasize on generating prediction regions with predefined coverage rates. This helps the decision-maker to know in advance what  the degree of constraint violation is upon obtaining the solution of the  optimization problem. Sharpness relates to how small  the spread of uncertainty is for the required probability guarantee. Too large prediction regions  increase cognitive load.   

In~\cite{golestaneh2018ellipsoidal}, we proposed a framework  to produce skilled ellipsoidal prediction regions. However, various decision-making problems demand for different forms of  uncertainty characterization. For example, the robust counterpart of a linear programming problem with polyhedral uncertainty sets is a linear programming problem while the same with the ellipsoidal uncertainty sets is a  Second Order Cone Programming (SOCP) problem~\cite{pachamanova2002robust}. Although, SOCP problems are convex and computationally tractable, their nonlinearity can be a practical drawback. 
Consequently, in this work we focus on multivariate prediction polyhedra. Our underlying motivation is to propose a data-driven approach   capable of generating highly skilled prediction polyhedra. We study evaluation methodologies for verification of the proposed methods using real data. Two formulations for prediction polyhedra are developed. In addition, due to recent interests in prediction convex hulls~\cite{soares2017active}, their relevance and limitations are discussed and a verification framework for their quantitative assessment  is developed. Because any forms of multivariate prediction is prone to be affected by outliers, we propose an idea to make convex hulls more robust to outliers. The robustness of the prediction regions to outliers is also examined and compared. All  techniques output convex polyhedra and suit the requirements of robust and chance-constrained optimization.
 Also, theoretically they can be employed for both spatial and temporal uncertainty prediction. Their performances in practice, however, will be assessed over empirical results in Section \ref{Section:Results}. The efficiency of the proposed frameworks is evaluated for wind and PV power. Temporal and spatial prediction polyhedra of dimensions 2, 3, 6, 12 and 24 with the probability of 5\% to 95\% in 5\% increments are generated and evaluated.
 
{The rest of the paper is organized as follows: in Section \ref{Methodology},  the proposed methodology and formulations to generate prediction polyhedra are discussed. The proposed skill assessment techniques  are provided in Section \ref{Section: Assessment}. The framework to estimate the parameters of the proposed prediction regions is explained in Section \ref{Section: Parameter Estimation}.  Section \ref{Section:Results} contains the empirical results and finally concluding remarks are given in Section \ref{Section:Conclusion}.}
\vspace{-0.7em}
\section{Methodology}
\label{Methodology}
 Due to  growing interest in characterizing  uncertainty information in  forms of polyhedra and multivariate boxes, in this section,  four  frameworks to produce such prediction geometries are  proposed. %
\vspace{-0.6em}
\subsection{Simple Prediction Polyhedra}
\label{Subsection:MIT polyhedra}
At every time step $t$, one aims at predicting the random variable, e.g. wind/PV power, for future times $t+1$, $t+2$,$...$, $t+K$ at $Z$ contiguous locations. Denote \textbf{X} as an uncertain variable of dimension $D=K \times Z$, $\textbf{X}_t=[X_{t+1},...,X_{t+D}]$. Denote $ \mu= E(\textbf{X})$ as the expected value \textbf{X} and   $ \varSigma= E[(\textbf{X}-\mu)(\textbf{X}-\mu)^\top] $ as its the covariance matrix. Inspired by~\cite{pachamanova2002robust}, we propose the following two formulations  for  prediction polyhedra.  
 \begin{equation}
\label{Eq:polyhedralgenralp1}
 P^1_{t,\alpha}:=\{\textbf{X} \ | \ \norm {\Lambda_t(\textbf{x}_t-\mu_t)^\top}_1\leq \Gamma_t^\alpha\}\\
\end{equation}
 \begin{equation}
\label{Eq:polyhedralgenralp2}
P^{\infty}_{t,\alpha}:=\{\textbf{X} \ | \ \norm {\Lambda_t(\textbf{x}_t-\mu_t)^\top}_\infty\leq \Delta_t^\alpha\}\\
\end{equation}
 where $\alpha$ is  the nominal coverage rate of prediction polyhedra, $\Delta$ and $\Gamma$ are called scale or robust parameters. With the assumption that $\varSigma^{-1}$ is a symmetric and positive definite matrix, $\Lambda$ as the Cholesky decomposition of $\varSigma^{-1}$ is an upper triangular matrix with positive diagonal elements. For  vector $\textbf{x}\in\mathbb{R}^{D\times1}$, $\norm{\textbf{x}}_1$ denotes the first norm  as $\sqrt{\sum_{d=2}^{D}|x_d|}$ and $\norm{\textbf{x}}_\infty$ denotes the infinity norm given by $\max _{d=1,...,D}|x_d|$. Henceforth, the upper case letters symbolize random variables while lower case letters express their realizations. 
 
 The polyhedron given by \eqref{Eq:polyhedralgenralp1} is  inscribed in a ellipsoid defined by the following formulation.
  \begin{equation}
 \label{Eq:EllipsosoidTypical}
   E_{t,\alpha}:= (\textbf{x}_t-\mu_t)^\top \varSigma^{-1}(\textbf{x}_t-\mu_t)\leq ({\Gamma_t^\alpha})^2\\
 \end{equation}
 
 The predictive performance of the uncertainty sets in forms $P^1$, $P^\infty$ is directly linked to how accurate and optimal their  predicted parameters are. The parameters to be predicted include a location parameter $\mu$ (mean vector), a shaping parameter $\varSigma$ (covariance matrix), and  scaling parameters $\Gamma^\alpha$ and $\Delta^\alpha$ (being a function of the nominal coverage rate). It is worth noting that even though the first and the second-order moment information (i.e., mean and covariance)  are classically used for Gaussian objects, considering them as a basis for defining polyhedra does not necessarily means one can assume the underlining distribution is Gaussian.
 
 In robust optimization literature, the scale parameter is commonly known as the uncertainty budget and it controls the conservativeness. The  uncertainty budget is determined by the user arbitrarily  based on his aversion to uncertainty. One does not expect to get uncertainty sets with predefined probability levels based on the common approaches available for determination of the uncertainty budget~\cite{li2015modeling,pachamanova2002robust}.
 
\begin{figure}[!t]
 	\vspace{-0.4em}
 	\centering
 	\includegraphics[width=7cm,height=6cm]{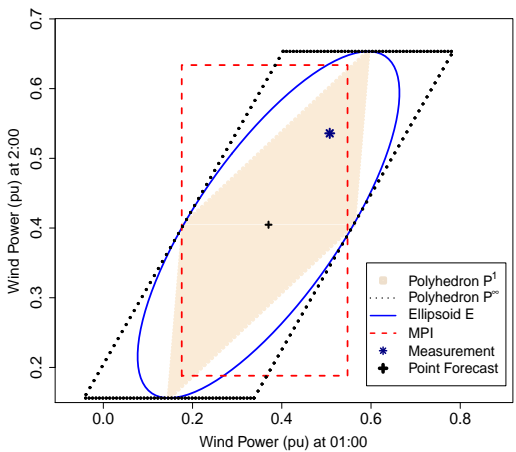}
 	\caption{Typical prediction geometries of dimension two.}
 	\label{Fig:Typical}
 	\vspace{-0.4em}
 \end{figure}

 Assuming equal values for $\Gamma$ and $\Delta$ in \eqref{Eq:polyhedralgenralp1}-\eqref{Eq:EllipsosoidTypical}, typical
 $P^1$, $P^\infty$ and $E$ with probability level of 85\% are illustrated in Fig. \ref{Fig:Typical}. In Fig. \ref{Fig:Typical} the predicted scale parameter is $2.210$ while the location $\mu$, and shape $\varSigma$ for $P^1$, $P^\infty$ and $E$ are
  \begin{equation}
 \label{Eq:Typical Pinf parameters}
 \varSigma=
 \begin{bmatrix}
 0.01762222 &  0.01135601 \\
 0.01135601 & 0.01265258 
 \end{bmatrix}, \ \ \mu=\begin{bmatrix}
 0.370 & 0.405  
 \end{bmatrix}\\
 \end{equation}
 
 It is to be emphasized that $\Gamma$ and $\Delta$ are not expected to be equal in general.  In Section \ref{Section: Parameter Estimation}, the proposed ideas to determine the correct values for $\Gamma$ and $\Delta$  are explained. What Fig. \ref{Fig:Typical}  illustrates  is that in case $\Gamma$ and $\Delta$ take equal values in \eqref{Eq:polyhedralgenralp1}-\eqref{Eq:EllipsosoidTypical}, how $P^1$, $P^\infty$ and $E$ relate to one another. Robust or interval optimization go along the faces/edges to find the optimal solution. From Fig. \ref{Fig:Typical}, it can be inferred that $P^1$ and $P^\infty$ impose similar computational cost in optimization because they actually have equal number of edges/facets with a difference that $P^1$ tends to be sharper. As shown in Fig. \ref{Fig:Typical}, the measurement is included in all $P^1$, $P^\infty$ and $E$. If having many more sample observations, one would expect close to 85\% of the observations to be covered by the prediction regions.
 \vspace{-0.7em}
\subsection{Prediction Convex Hulls}
 \label{Subsection:Convex Hulls formulation}
\begin{figure}[!t]
	\vspace{-0.4em}
	\centering
	\includegraphics[width=8cm,height=7cm]{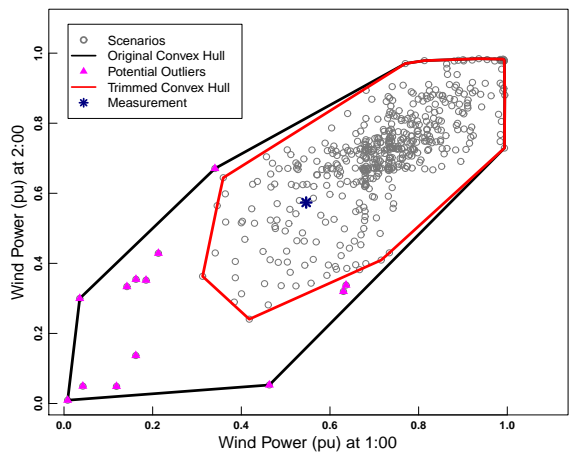}
	\caption{{The convex hull containing all the predicted scenarios. The original convex hull against the trimmed convex hull generated by excluding the potential outliers
			.}}
	\label{Fig:TrimmedHull1}
	\vspace{-0.4em}
\end{figure}

The convex hull of a set  of points, $S$, is the smallest convex set  containing all the points. The idea  is to find the convex hull of spatial/temporal scenarios~\cite{golestaneh2016generation}. Spatial/temporal scenarios are generated by uniformly  sampling from multivariate predictive distributions. At each time $t$, $S$ scenarios are produced where each scenario is a vector of dimension $D$.  Among  few methods available to find a non-ambiguous and efficient representation of  required convex hulls, we use Quickhull algorithm. Quickhull algorithm is  fast and efficient in most cases and tends to perform well in practice~\cite{barber1996quickhull}. Time complexity of this algorithm for most cases is $O(n \log n)$ and in the worst case is $O(n^2)$. Theoretically, Quickhull algorithm can work in high dimensions. For a straightforward explanation and implementation guide of Quickhull algorithm, one can refer to~\cite{mucke2009quickhull}. The original convex hull illustrated in Fig. \ref{Fig:TrimmedHull1} represents the convex hull of predicted temporal scenarios of wind power for a randomly selected day.
\vspace{-0.4em}
\subsection{Trimmed Prediction Convex Hulls}
\label{Subsection: Outliers}
 As stated in subsection \ref{Subsection:Convex Hulls formulation}, the predicted spatial/temporal scenarios are produced by uniformly sampling from  multivariate predictive distributions. Although in the Monte Carlo-based analysis, all scenarios are considered to come with an equal likelihood of occurrence, some of them might be far from the center of cloud. We label those scenarios 
  as outliers. Outliers are marked in Fig.  \ref{Fig:TrimmedHull1}. As can be observed, they grossly impact on the size of prediction convex hulls.  Discarding the outliers results in the trimmed convex hull in  Fig.  \ref{Fig:TrimmedHull1} which is much sharper than the original one.

Although there is a wealth of techniques available to detect outliers in univariate datasets,  only limited options are at hand for multivariate data. It is to be noted that a multivariate outlier does not necessarily have to be an outlier in any of its univariate coordinates. Mahalanobis distance is one of the most widely used metrics for multivariate outlier identification. Basically, it quantifies how far away a point is from the center of the cloud, taking into account the shape of the cloud as well. Those scenarios with Mahalanobis distance larger than the critical chi-square  values  at a significance level of 0.001 are labeled as outliers~\cite{werner2003identification}.
\vspace{-0.4em}
 \subsection{Benchmark Method}
 \label{subsection:MPI formulation}
 The multivariate prediction literature still is in a primitive stage and there are not many data-driven benchmarks available to conduct a comparative study on   the performance of the proposed techniques. Among few works available, the adjusted intervals approach is found to be a relevant  benchmark~\cite{bessa2015marginal, kolsrud2007time, li2011simultaneous}. 
 This technique  uses the marginal (univariate) prediction intervals and the  multivariate scenarios as the inputs to generate Multivariate Prediction Intervals (MPIs). For the approach to generate MPIs, the reader is referred to~\cite{bessa2015marginal}. Typical  MPIs are illustrated in Fig. \ref{Fig:Typical}.
 
\vspace{-0.6em}
 \section{Predictive Skill Assessment}
 \label{Section: Assessment}
 The predictive performance of probabilistic forecasts are commonly examined based on their two properties, namely calibration and sharpness. Calibration is a joint property of forecasts and observations, and  it is decided based on the statistical consistency between  them. Sharpness refers to concentration of forecasts~\cite{gneiting2007probabilistic}. Following the probability and statistics literature,  we refer to calibration as the proximity of the nominal coverage rate of a prediction region to its empirical coverage rate. The coverage rate of a prediction region is the proportion of times that the region contains materialized events (observations). Similarly, a nominal coverage rate refers to the expected coverage while an empirical coverage represents the empirical coverage of that region calculated based on real data. Sharpness is examined based on the size of prediction regions, e.g. area in dimension two, and volume in higher dimensions. The aim  is to generate sharp and concentrated prediction polyhedra subjected to calibration.
 \vspace{-0.6em}
 \subsection{Simple Prediction Polyhedra}
  As can be observed in Fig. \ref{Fig:Typical}, both $P^1$ and $P^\infty$ are simple, convex and have few edges. In geometry, based on the definition, each vertex of a $D$-dimensional simple polyhedron is adjacent to exactly $D$ edges. Robust and interval optimization go along  the edges of uncertainty sets to find the optimal solution. In general, fewer number of faces  is an advantage in the sense that it imposes less computation to optimization.
 
 There is no limitation to represent uncertainty in higher dimensions in the form of $P^1$ and $P^\infty$ as long as the correlation matrix $\varSigma$ can be predicted and Cholesky decomposition of $\varSigma^{-1}$ can be calculated.  The proposed approach  and formulations are competent at generating prediction polyhedra with any desired probability guarantees.  
 
 \textbf{Volume}: Since to the best of our knowledge there is no straightforward approach to calculate the volume of $P^1$ and $P^\infty$ analytically, in Section \ref{Section:Results}, a Monte Carlo-based technique is explained to estimate their volumes numerically. 
 
 \textbf{Calibration}: To evaluate calibration of prediction polyhedra, one needs to calculate the empirical coverage of each predicted polyhedron and compares that with the corresponding nominal coverage. Let $\xi_{t}^{\alpha_i}$ be a binary  variable taking 1  if  the prediction polyhedron with nominal probability $\alpha_i$ contains the observed value at time $t$ and 0 otherwise. Then the empirical coverage is given by
 \begin{equation}
 \label{Eq:calib}
 \hat{\alpha}_i = \frac{1}{T} \sum_{t=1}^{T} {\xi_{t}^{\alpha_i}}
 \end{equation}
 with $T$ as the length of the evaluation set. $P^1$ and $P^\infty$ include the measurement $\textbf{x}_t$ if it satisfies \eqref{Eq:polyhedralgenralp1} and  \eqref{Eq:polyhedralgenralp2}, respectively.
 \vspace{-0.6em}
 \subsection{Prediction Convex Hulls}
\begin{figure}[!t]
 	\vspace{-0.4em}
 	\centering
 	\includegraphics[width=7cm,height=6cm]{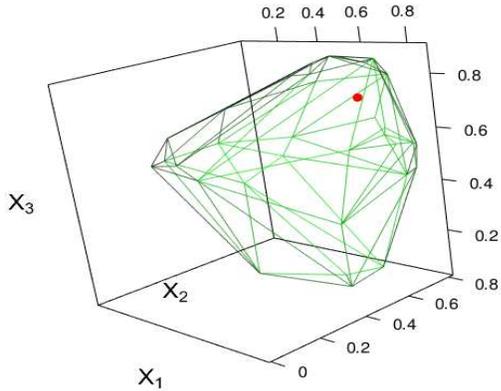}
 	\caption{{A typical convex hull generated for PV power data. $X_1$, $X_2$ and $X_3$ denote PV power output at zones 1, 2 and 3, respectively. The red dot represents the measurement.}}
 	\label{Fig:PV3D}
 	\vspace{-0.4em}
 \end{figure}

 As one can notice in Fig. \ref{Fig:TrimmedHull1}, the number of faces in convex hulls is higher than simple prediction polyhedra. This is more noticeable in higher dimensions as shown in Fig. \ref{Fig:PV3D}. A higher number of faces/edges imposes a higher computation into optimization because as mentioned before robust or interval optimization go along the faces/edges to find the optimal solution. In addition, a major limitation of  prediction convex hulls is that they cannot be generated for  predefined nominal coverage rates. They just represent the smallest convex region  including the predicted scenarios. When verifying them  on real measurements, they can show any empirical coverage, ranging between zero to one. Prediction convex hulls comparing to  $P^1$ and $P^\infty$ have the complexity of generating multivariate scenarios as their input first.

 \textbf{Volume}: One advantage of convex polyhedra is that their volumes can be calculated by subdividing them into smaller pieces. To do that a common approach is by triangulation methods where the polyhedron is decomposed into simplices. A simplex is a generalization of triangle to arbitrary dimensions. The volume of simplices can easily be computed. Eventually, the volume of polyhedron is computed by summing up the volumes of all simplices~\cite{bueler2000exact,barber1996quickhull}.
 
 \textbf{Calibration}: In convex geometry, given points $C=\{\hat{\textbf{x}}_1, \hat{\textbf{x}}_2, ..., \hat{\textbf{x}}_S\}$, the point $ \theta_1 \hat{\textbf{x}}_1+\theta_2 \hat{\textbf{x}}_2, ..., \theta_S \hat{\textbf{x}}_S $ is called their convex combination if  $ \theta_i\geq 0, i= 1,2, ..., S$ and $\sum_{i=1}^S \theta_i=1$. Therefore, a convex combination of points can be viewed  as a weighted average of the points, with $\theta$ as the weight of each point in the mixture. The convex hull of set $C$ ($\textbf{conv} \:C $) contains the arbitrary point $\textbf{y}$, if $\textbf{y}$ is a convex combination of $C$~\cite{boyd2004convex}. We use this definition to identify if a prediction convex hull generated for time $t$ includes the measurement recorded at the same time. $\textbf{y}\in C$ if there   is a solution for the following linear programming problem
 \begin{equation}
 \label{Eq:objective}
 \arg_{\theta} \min e^\top\boldsymbol{\theta}
 \end{equation}
 subject to	
 \begin{equation}
 \begin{gathered}
 \label{Eq:constraints}
 A\boldsymbol{\theta}=y\\
 B\boldsymbol{\theta}=1\\
 \boldsymbol{\theta}\geq 0
 \end{gathered}
\end{equation}
 with $\textbf{y} \in \mathbb{R}^D$, $e \in \mathbb{R}^D$ arbitrary  cost vector, $B= (1,...,1) \in \mathbb{R}^S$  $A=(\textbf{x}_1, ..., \textbf{x}_S) \in \mathbb{R}^{D \times S}$ and $\boldsymbol{\theta} \in \mathbb{R}^S$. 
 
 Following \eqref{Eq:objective} and \eqref{Eq:constraints}, one can determine if an arbitrary point $\textbf{y}$ is inside $\textbf{conv} \:C $ directly with no need to  generate $\textbf{conv} \:C $ first. The observed coverage of prediction convex hulls can be computed according to \eqref{Eq:calib}  once the inclusion of each $ \textbf{y}$ in the evaluation set is examined. 
 \vspace{-0.6em}
\subsection{Multivariate Prediction Intervals}
\textbf{Calibration}: For MPIs, the empirical coverage rate is computed by counting the number of measured  scenarios which fully lie within their boundaries~\cite{bessa2015marginal}.

\textbf{Volume}: The volume $V_t$  of a MPI at time $t$ is calculated as
 \begin{equation}
 \label{Eq:Volume_MPI}
 V_t=\prod_i(h_{i,t}-l_{i,t}) \quad\forall t
 \end{equation}
 with  $h_{i,t}$ and $l_{i,t}$ as the upper and lower
 bounds of the MPI at  dimension $i$, respectively.
 \vspace{-0.4em}
\section{Parameter Estimation}
\label{Section: Parameter Estimation}
The parameters of $P^1$ and $P^\infty$ are determined as

\boldsymbol{$u_{t}$}: The location parameter $u_{t}$  is the center of prediction polyhedra and is considered to be the point forecasts for the multivariate random variable $\textbf{X}$ at time $t$. Denote $\hat{\textbf{x}}_t=[\hat{x}_{1,t}\ \ \hat{x}_{2,t}\ \ ...\ \ \hat{x}_{D,t}]	 $, with $ \hat{x}_{i,t}, \: \forall i $  as the point forecast for time $ t $ and dimension $ i $ where  $ \hat{x}_{i,t} $ for each dimension is generated independently. We refer to  $ \hat{\textbf{x}} $ as predictions  and $ \textbf{x} $ as the measurement or materialized trajectory.

\boldsymbol{$\varSigma_t$}:
The $\varSigma_t$ is defined as the covariance matrix of point forecast errors estimated using  data up to time $t$. We suggest to use the Dynamic
Conditional-Correlation-GARCH (GARCH-DCC) technique to predict the covariance matrix~\cite{engle2002dynamic}.  In econometrics literature,  GARCH-DCC  has been widely implemented and is shown to be capable of estimating time-varying covariance matrices~\cite{engle2002dynamic,engle2001theoretical}. GARCH-DCC is mostly suitable for those random processes like forecast errors of renewable power generation for which the covariance matrix changes noticeably over time. In case the random process presents a slow-moving covariance matrix, rolling historical correlations and exponential smoothing as  less complicated techniques can be deployed~\cite{zakamulin2015test}.

\boldsymbol{$\Gamma_t^\alpha$}:
In~\cite{golestaneh2018ellipsoidal}, for ellipsoidal prediction regions, we proposed an approach to find the optimal scale parameters by making a compromise between volume of the ellipsoids and their calibration. To the best of our knowledge, for $P^1$ and $P^\infty$ polyhedra, there is no straightforward closed form formulation to calculate the volume. Therefore, we propose a data-driven technique  to find the minimum scale parameter which provides the required coverage rate over the most recent historical data. The scale parameter is updated whenever new measurements are received.  In the  proposed method, a window of size $\omega$ of the most recent measurements, point forecasts and predicted Cholesky decomposition of covariance matrices are input in the following equation and $\Upsilon$ for those values is calculated.
\begin{equation}
\label{Eq:GammaP1}
\Upsilon_i=\{\textbf{X}|\norm {\Lambda_i(\textbf{x}_i-\mu_i)^\top}_1 \} \qquad i=t-\omega,...,t-2,t-1\\
\end{equation}
where $\mu_i$ is $\hat{\textbf{x}}_i$ and $\textbf{x}_i \: \forall i$ are the measured trajectories. Then, $\Upsilon_i \:, \forall i$ are sorted ascending. For the desired probability level $\alpha$, $\Gamma_t^\alpha$ is considered as the $N^{th}$ smallest $\Upsilon_i, \: \forall i$ or in other words the $N^{th}$  element of the sorted vector, where $N$ is 
\begin{equation}
\label{Eq:GammaP1_round}
N=\text{round} (\omega\times\alpha)\\
\end{equation}
with round($x$) as a function which returns the closest integer to $x$. Following the proposed method, $\Gamma_t^\alpha$ is updated for each $t$ on a rolling base. This technique is based on this expectation that if prediction polyhedra envelop a window of most recent historical data, they  should present a similar coverage for the  future observations. After obtaining $u_{t}$, $\varSigma_t$ and $\Gamma_t^\alpha$, $P^1_{t,\alpha}$ is readily available.

\boldsymbol{$\Delta_t^\alpha$}:
The $\Delta_t^\alpha$ can be estimated similar to $\Gamma_t^\alpha$ as explained above, with the only difference that $\Upsilon_i \:, \forall i$ are calculated as
\begin{equation}
\label{Eq:GammaPinf}
\Upsilon_i=\{\textbf{X}|\norm {\Lambda_i(\textbf{x}_i-\mu_i)^\top}_\infty \} \qquad i=t-\omega,...,t-2,t-1\\
\end{equation}
 
For techniques to generate spatial/temporal scenarios, MPIs and convex hulls, the reader is referred to~\cite{golestaneh2016generation},~\cite{bessa2015marginal} and~\cite{mucke2009quickhull}, respectively.
\vspace{-0.6em}
\section{Empirical Results}
\label{Section:Results}
In order to evaluate the proposed forecasting frameworks, the wind power and PV power datasets provided for the Global Energy Forecasting Competition (GEFCom) 2014 are used here. The datasets are available online~\cite{website55}. We use wind power data to predict temporal dependency and PV power data  to study  spatial dependency.   The wind power dataset includes  wind power measurements of 10 wind farms in Australia. The data for farm three is used here for analysis. The data includes four explanatory variables which are zonal and meridional wind components  forecasts at two heights, 10 and 100 m above ground level provided by the European Centre for Medium-range Weather Forecasts (ECMWF). Weather forecasts are  issued every day at midnight. The resolution of data is of one hour and forecast horizons are 1- to 24-hour ahead. Data for January 2012 to the end of April 2013 is used to train the models while the out of sample subset covers May 2013 to December 2013. PV power data includes 12  independent  variables as the output of Numerical Weather Predictions (NWPs) used as predictors and PV power generation as predictand. The available data covers the period from April 2012 to the end of June 2014 for three contiguous  zones. Data for April 2012 to the end of May 2013 is used to train the model and the evaluation subset covers data from June 2013 to the end of May 2014. Analysis are carried out to predict the simultaneous stochastic behavior of PV power at  three zones at 12:00 pm for spatial dependency studies. Power measurements are normalized by the nominal capacity of their corresponding generation unit. 

A support vector machine (SVM)  from package ``e1071" in R whose parameters are tuned based on 5-fold cross-validation is used to generate wind/PV power point forecasts. It yields  15.43\%, 14.1\% root mean square error for wind and PV power (12:00 pm only), respectively.   Because all wind farms are adjacent to each other, the weather forecasts available for the first six wind farms are used as the explanatory variables to generate forecasts for farm 3. The covariance matrices are predicted using DCC-GARCH  functions from ``rmgarch" package in R. Univariate quantiles with the nominal probability 2.5\% to 97.5\% in 2.5\% increments  are produced by quantile regression. 
 500  scenarios~\cite{golestaneh2016generation} are generated as the inputs for adjusted interval technique. The upper limit of intervals is considered to be 99.5\% quantile given by quantile regression and the lower limit is considered to be zero.

\begin{figure*}[t!]
	\begin{tabular}[c]{ccc}
		\begin{subfigure}{.33\textwidth}
			\centering
			\includegraphics[width=\linewidth,height=5.3cm]{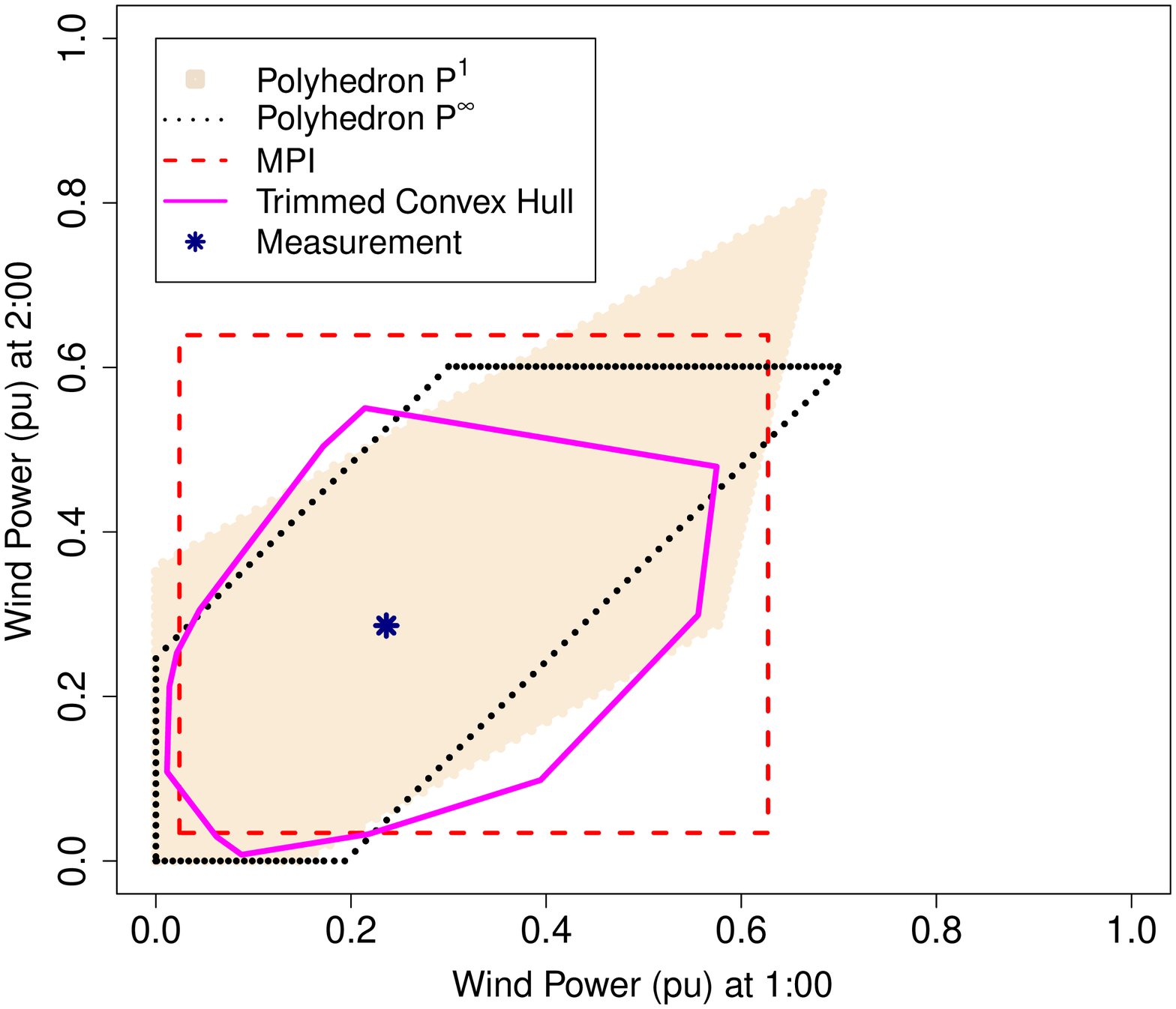}
			\label{fig:Sample Day 1}
		\end{subfigure}
		\begin{subfigure}{.33\textwidth}
			\centering
			\includegraphics[width=\linewidth,height=5.3cm]{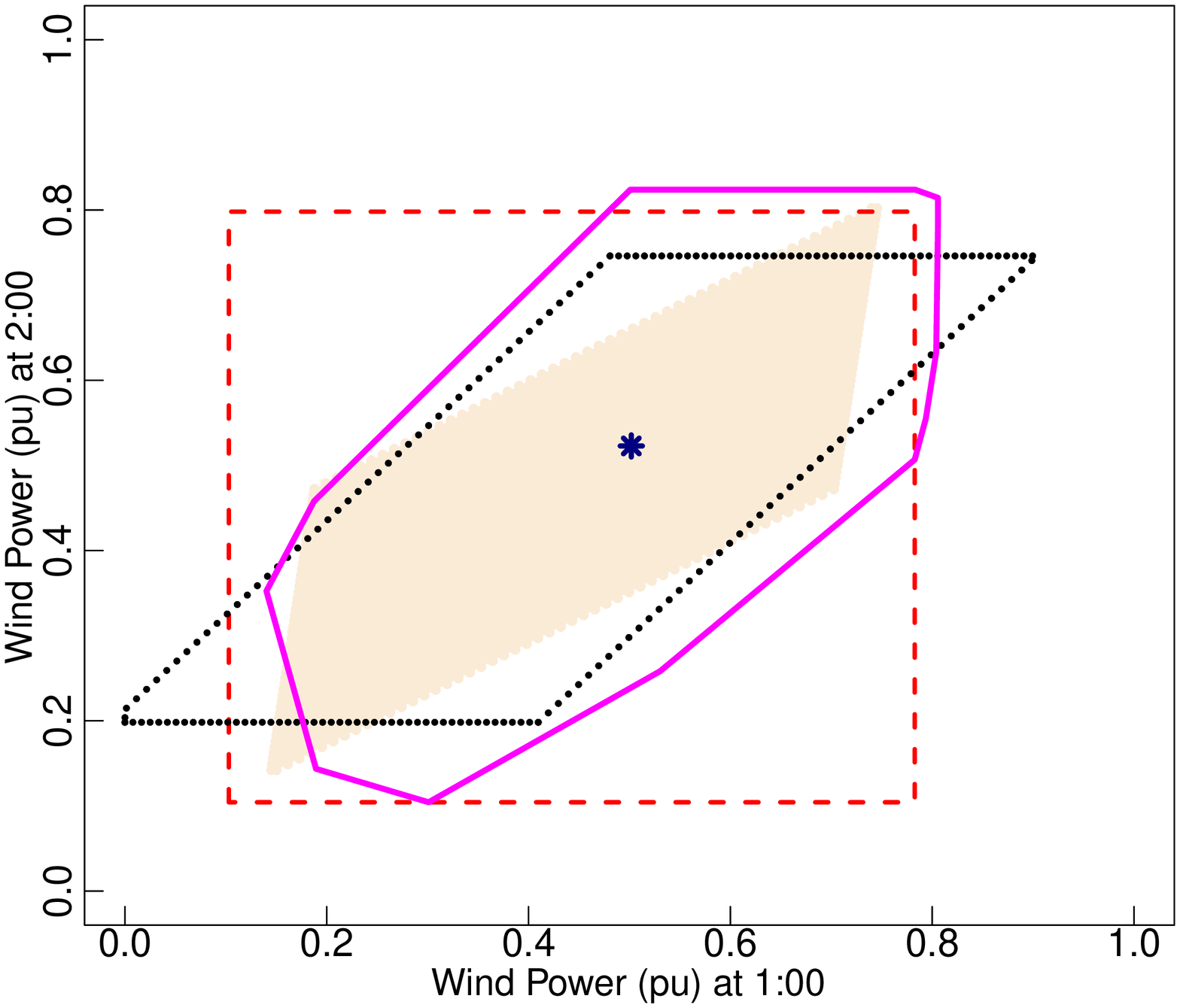}
			\label{fig:Sample Day 2}
		\end{subfigure}
		\begin{subfigure}{.33\textwidth}
			\centering
			\includegraphics[width=\linewidth,height=5.3cm]{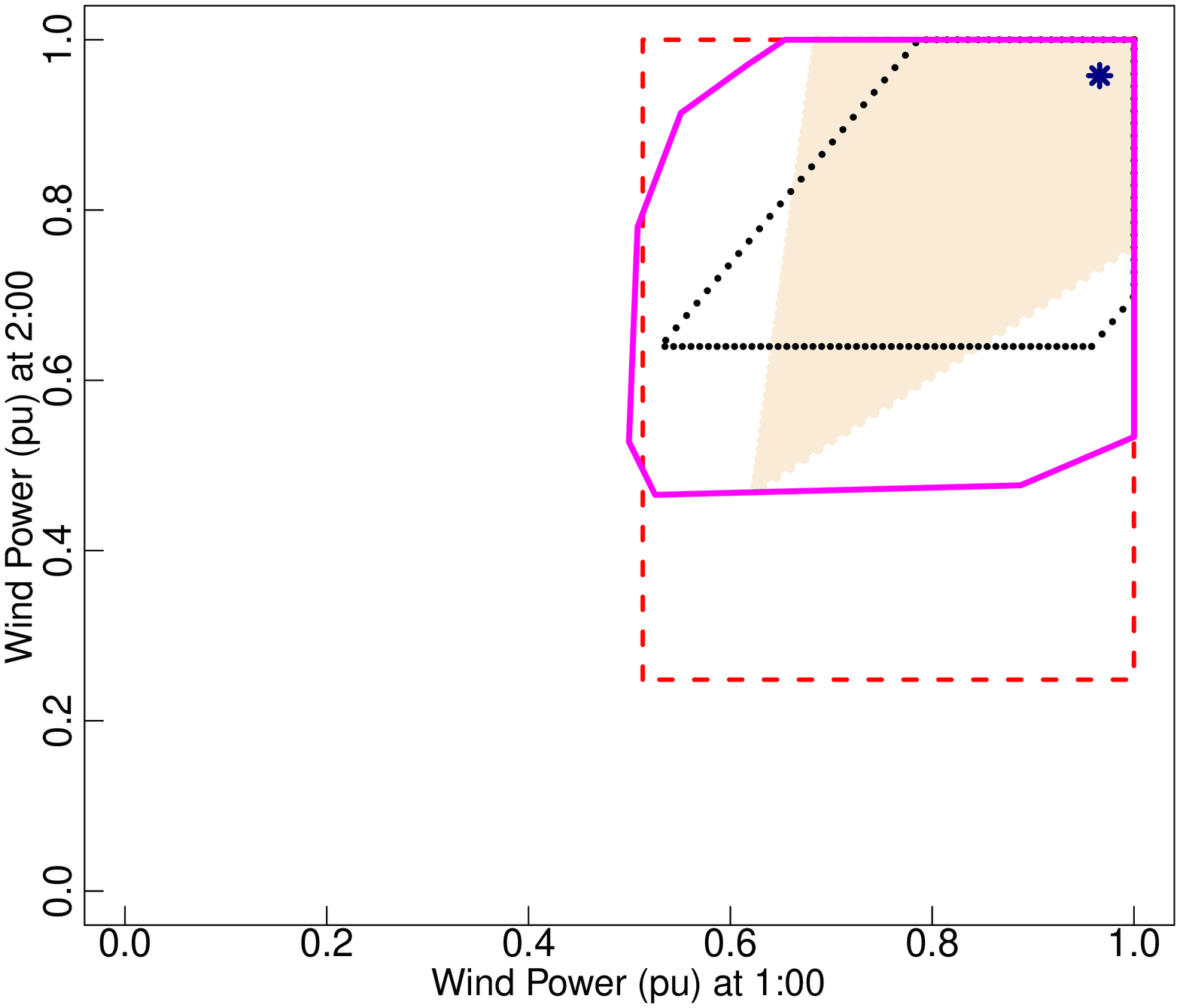}
			\label{fig:Sample Day 3}
		\end{subfigure}%
	\end{tabular}
	\caption{Visual comparison of temporal prediction polyhedra of dimension 2 for wind power data.  \label{fig:Sample Days}}
	\vspace{-0.5em}
\end{figure*}

Prediction polyhedra of dimensions 2, 6, 12 and 24 for wind power data are generated and evaluated. Dimension 2 includes wind power data at 01:00 am and 2:00 am. Dimension 6 covers 1- to 6-hour head predictions from 01:00 am to 6:00 am. Dimension 12 represents data from 01:00 am to 12:00 pm and dimension 24 includes all 24 hourly lead times from 01:00 am to 24:00 midnight. Prediction polyhedra produced for PV power data are of dimension 3, describing the correlated uncertainty of PV power at three zones under the study at 12:00 pm. Throughout this section, all the analysis in dimension 3 are based on spatial prediction polyhedra produced for PV power data while the results provided in other dimensions are based on temporal prediction polyhedra of wind power.

Fig. \ref{fig:Sample Days} shows the prediction polyhedra for three randomly selected days for out of sample data. The regions are limited to the feasible range of normalized wind power data [0,1]. Comparing MPIs with $P^1$,  $P^\infty$ and convex hulls, one can notice that the later ones present a correlated pattern between generation at two successive hours while MPIs show a uniform relation between them. All polyhedra have a fairly reasonable size and follow the variations in wind power generation.

In the following, we will compare the various prediction polyhedra in dimensions $D\geq 2$. The following simulations results suggest that both  $P^1$ and  $P^\infty$ have better predictive skill than MPIs in terms of both calibration and sharpness (volume). In addition, $P^1$ tends to be sharper and less conservative than $P^\infty$. It is to be noted that one should expect to see more improvements in the area of verification of such forecasts in the future. We still have a minimum sound basis here to analyze our forecasts and conclude.

\begin{figure*}[h!]
	\centering
	\begin{tabular}[c]{c}
		\begin{subfigure}{.45\textwidth}
			\centering
			\includegraphics[width=\linewidth,height=5.0cm]{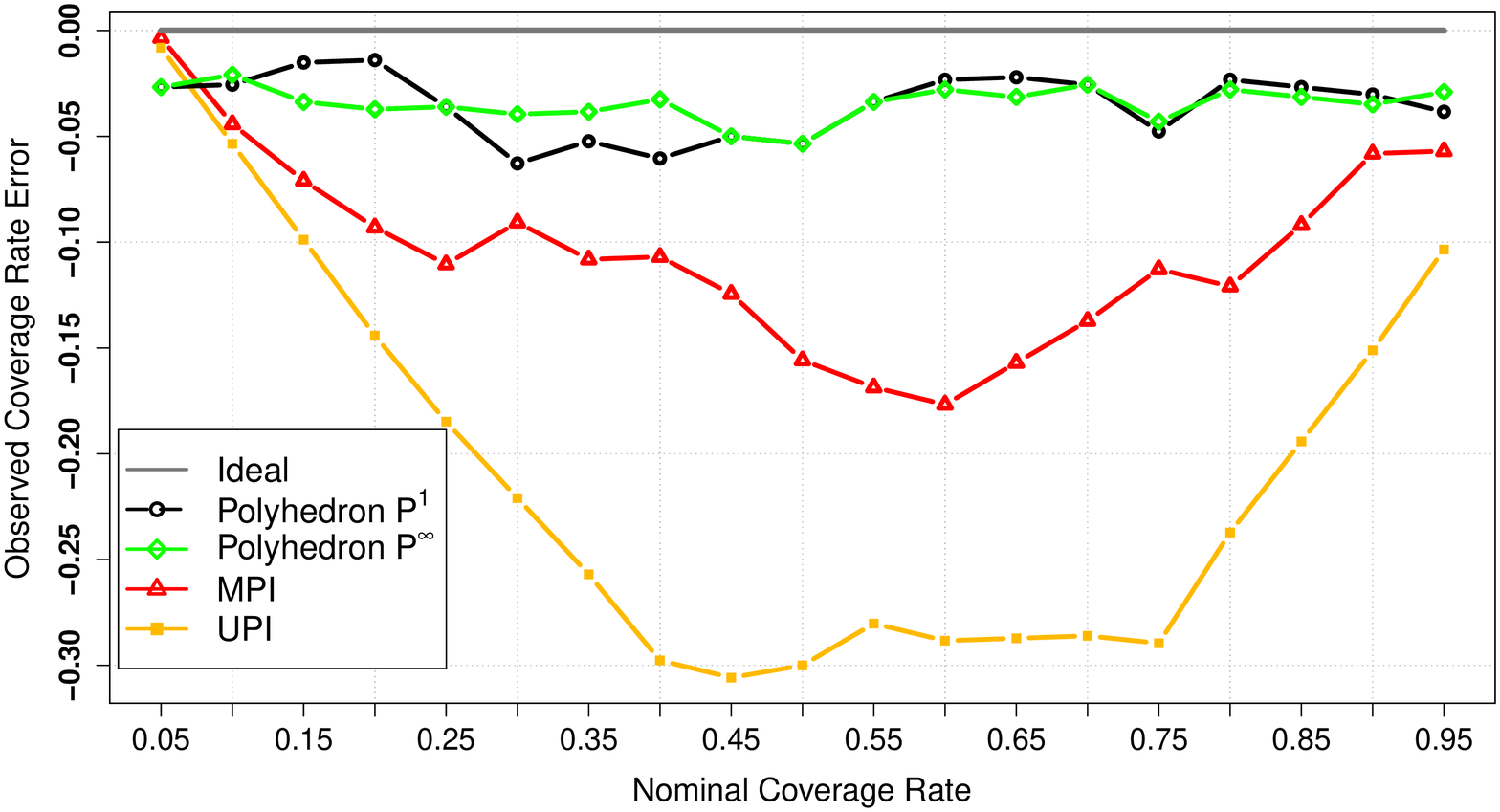}
			\caption{Temporal polyhedra, dimension 2}
			\label{fig:Cal Dim2}
		\end{subfigure}
		\begin{subfigure}{.45\textwidth}
			\centering
			\includegraphics[width=\linewidth,height=5.0cm]{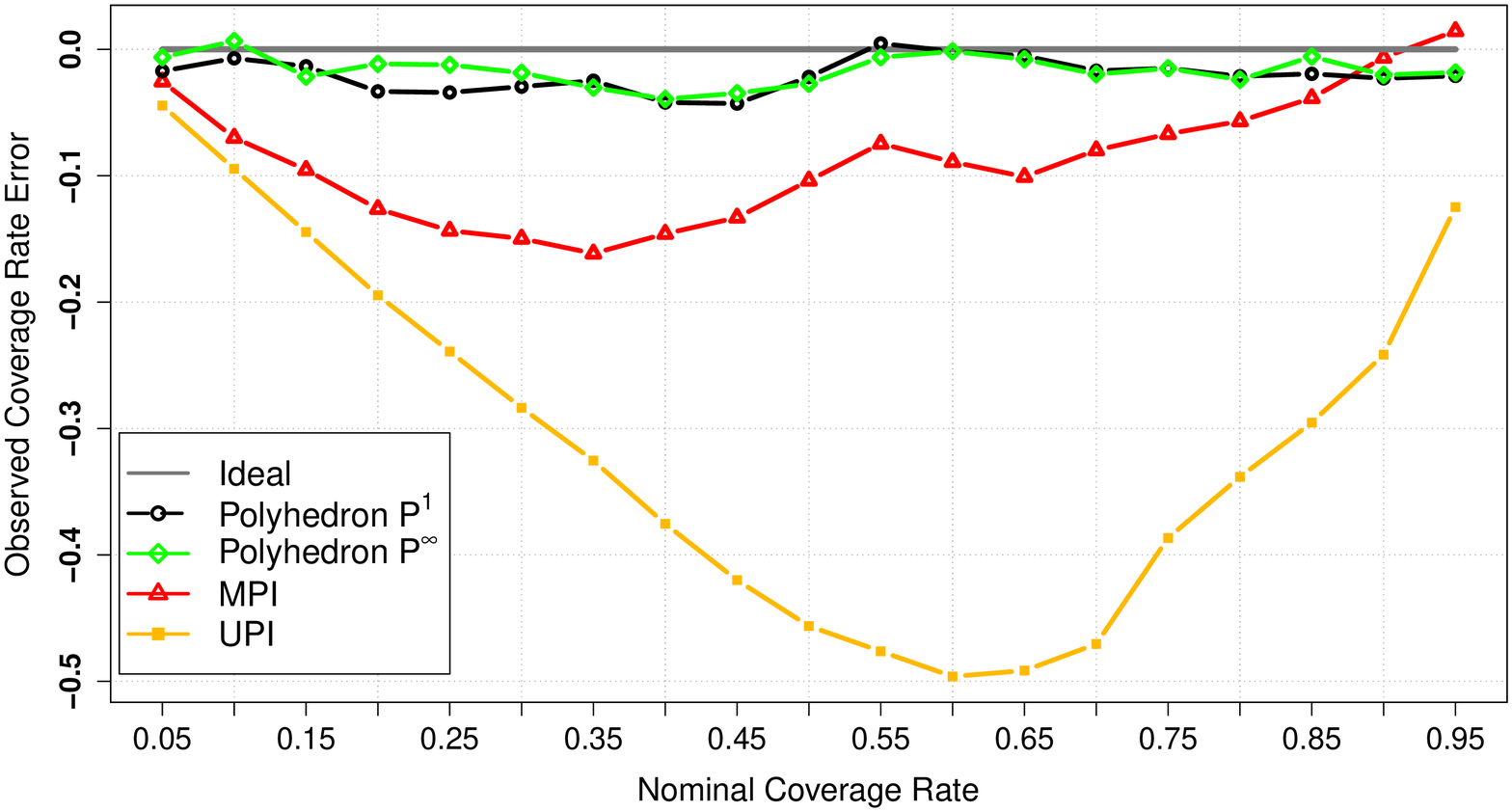}
			\caption{Spatial polyhedra, dimension 3}
			\label{fig:Cal Dim 3}
		\end{subfigure}
		\\
		\begin{subfigure}{.45\textwidth}
			\centering
			\includegraphics[width=\linewidth,height=5.0cm]{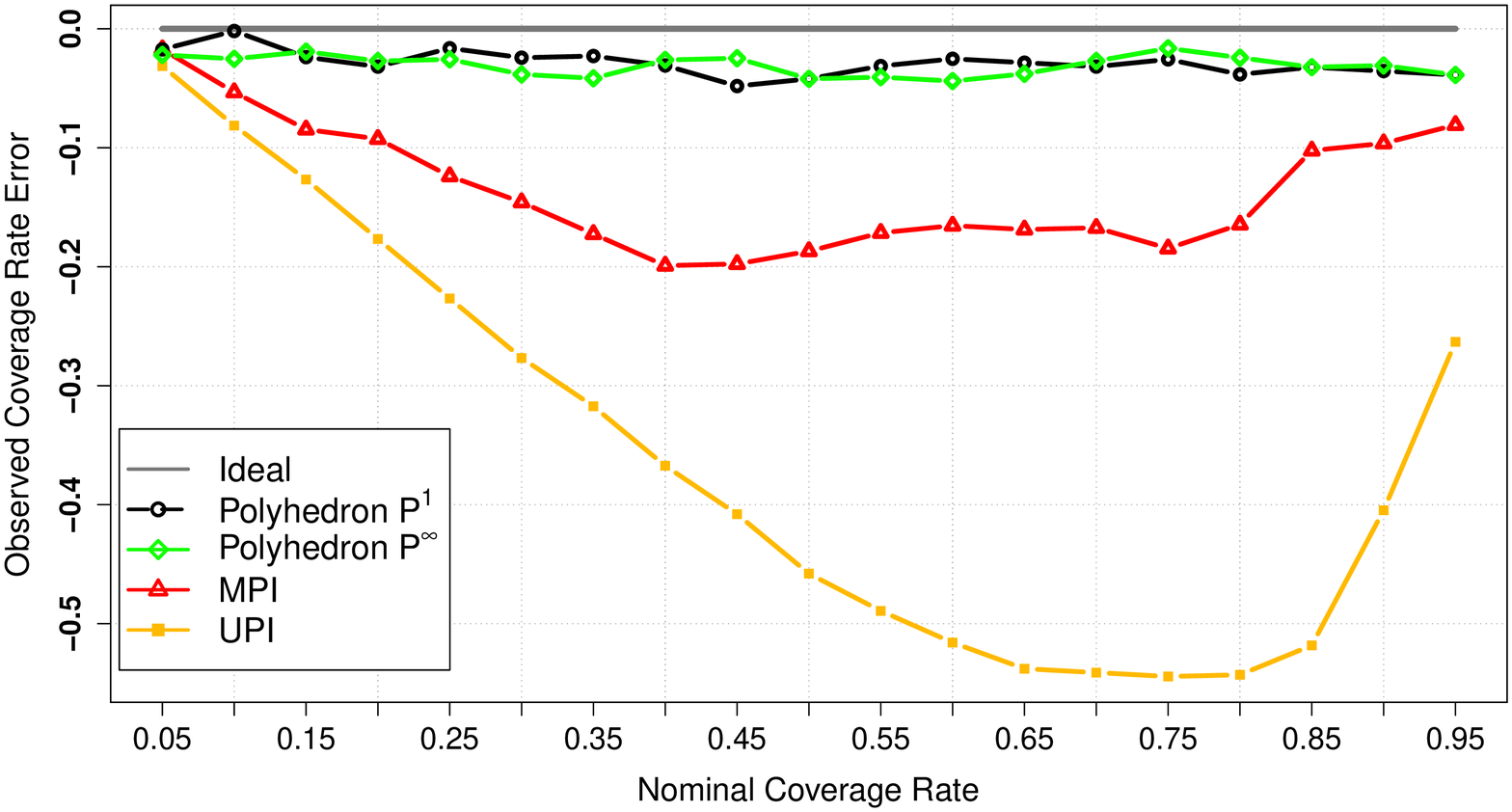}
			\caption{Temporal polyhedra, dimension 6}
			\label{fig:Cal Dim 6}
			\vspace{-0.6em}
		\end{subfigure}%
		\begin{subfigure}{.45\textwidth}
			\centering
			\includegraphics[width=\linewidth,height=5.0cm]{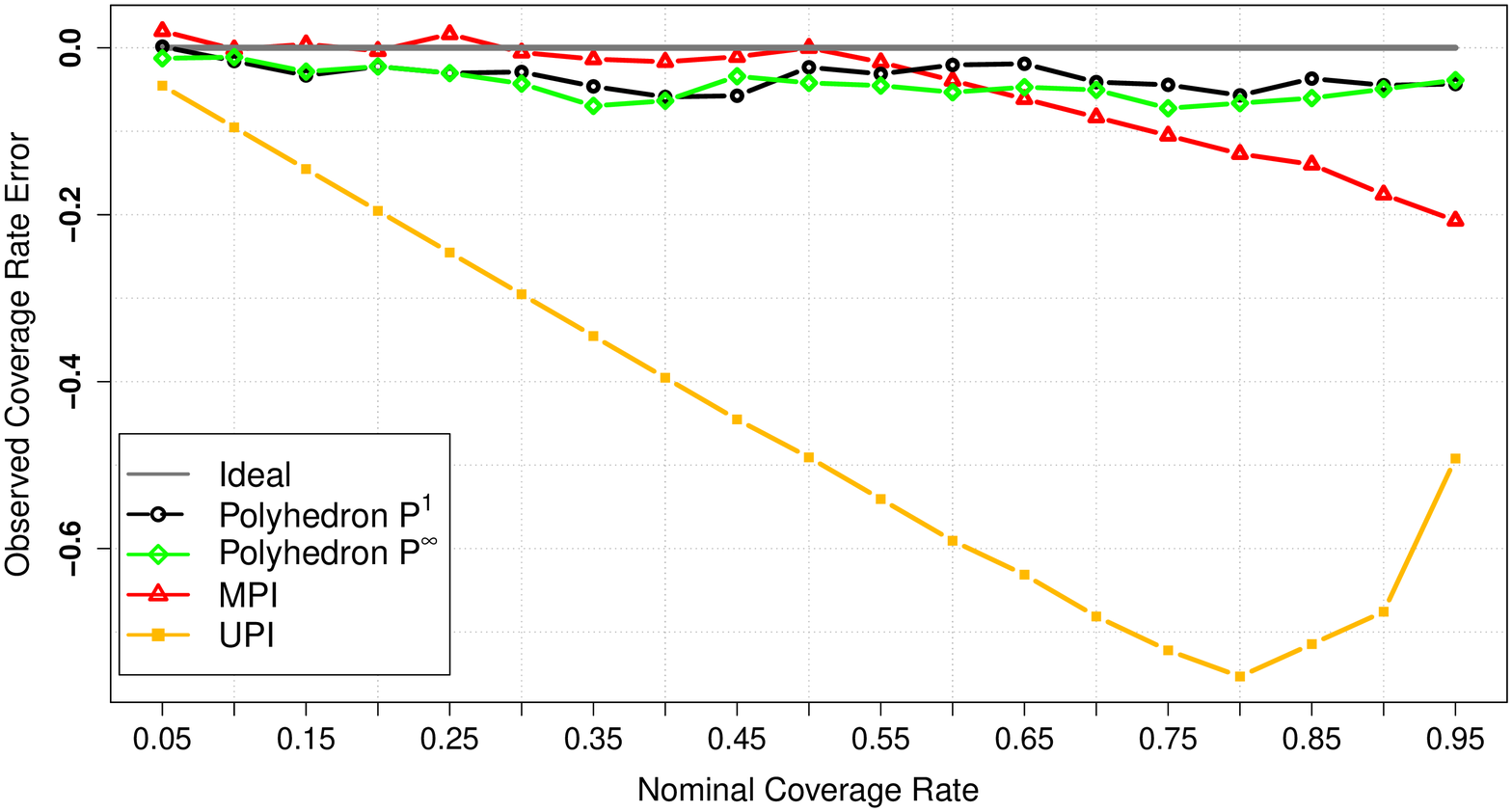}
			\caption{Temporal polyhedra, dimension 24}
			\label{fig:Cal Dim 24}
			\vspace{-0.6em}
		\end{subfigure}%
	\end{tabular}
	\caption{The difference between empirical coverage  and nominal coverage rates (Empirical-Nominal). \label{fig:Calibration Error}}
	\vspace{-0.5em}
\end{figure*}
Fig. \ref{fig:Calibration Error} reports the deviations between empirical coverage rate and nominal coverage rate of prediction polyhedra for dimensions 2, 3, 6 and 24. The nominal coverage rates ranging from 0.05 to 0.95 in 0.05 increments are included in the figure. As it is expected, UPIs  fail to capture the dependent and correlated uncertainty of wind/PV power output over successive hours and at adjacent locations. The calibration and reliability of UPIs decline as the dimension increases. The calibration of MPIs is also woefully inadequate. The  $P^1$ and $P^\infty$ polyhedra  maintain a fairly stable calibration in all dimensions and for all nominal coverage rates. Convex hulls are not covered in this figure because as discussed in Section \mbox{\ref{Section: Assessment}}, they do not provide prediction regions with predetermined nominal coverage rates. When using prediction convex hulls, one expects to get the smallest convex region with the highest probability guarantee. The untrimmed temporal  convex hulls return 90.4\%, 56\%, 10\% and almost 0\% empirical coverage rates in dimensions 2, 6, 12 and 24 for wind power, respectively. The spatial prediction convex hulls  contain 90\% of the PV power measurements.  Our empirical results suggest that the  prediction convex hulls perform poorly in higher dimensions. We produced temporal prediction convex hulls of dimension 4  for wind power data (1:00 am to 4:00 am) and obtained 72\% empirical coverage rate. Thus, we do not recommend prediction convex hulls in dimensions higher than 4. Additionally, although it is straightforward and computationally efficient to compute  calibration of convex hulls following \eqref{Eq:objective} and \eqref{Eq:constraints}, the algorithm to find  convex hulls themselves becomes very slow for dimensions higher than 8 and it does not converge in dimensions more than 9.

\begin{figure*}[h!]
	
	\begin{tabular}[c]{ccc}
		\begin{subfigure}{.33\textwidth}
			\centering
			\includegraphics[width=.9\linewidth]{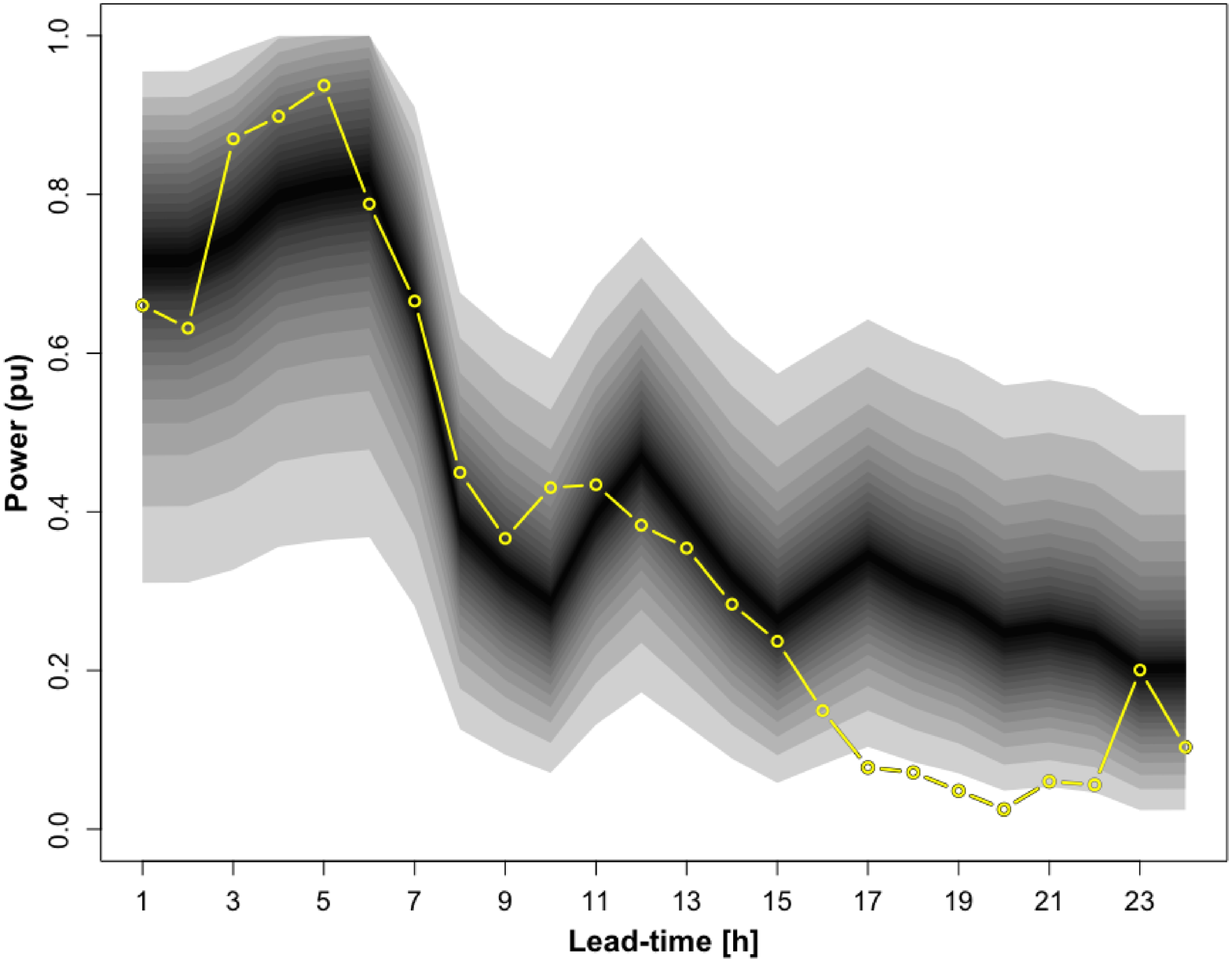}
			\caption{}
			\label{fig:UPI1}
		\end{subfigure}
		\begin{subfigure}{.33\textwidth}
			\centering
			\includegraphics[width=.9\linewidth]{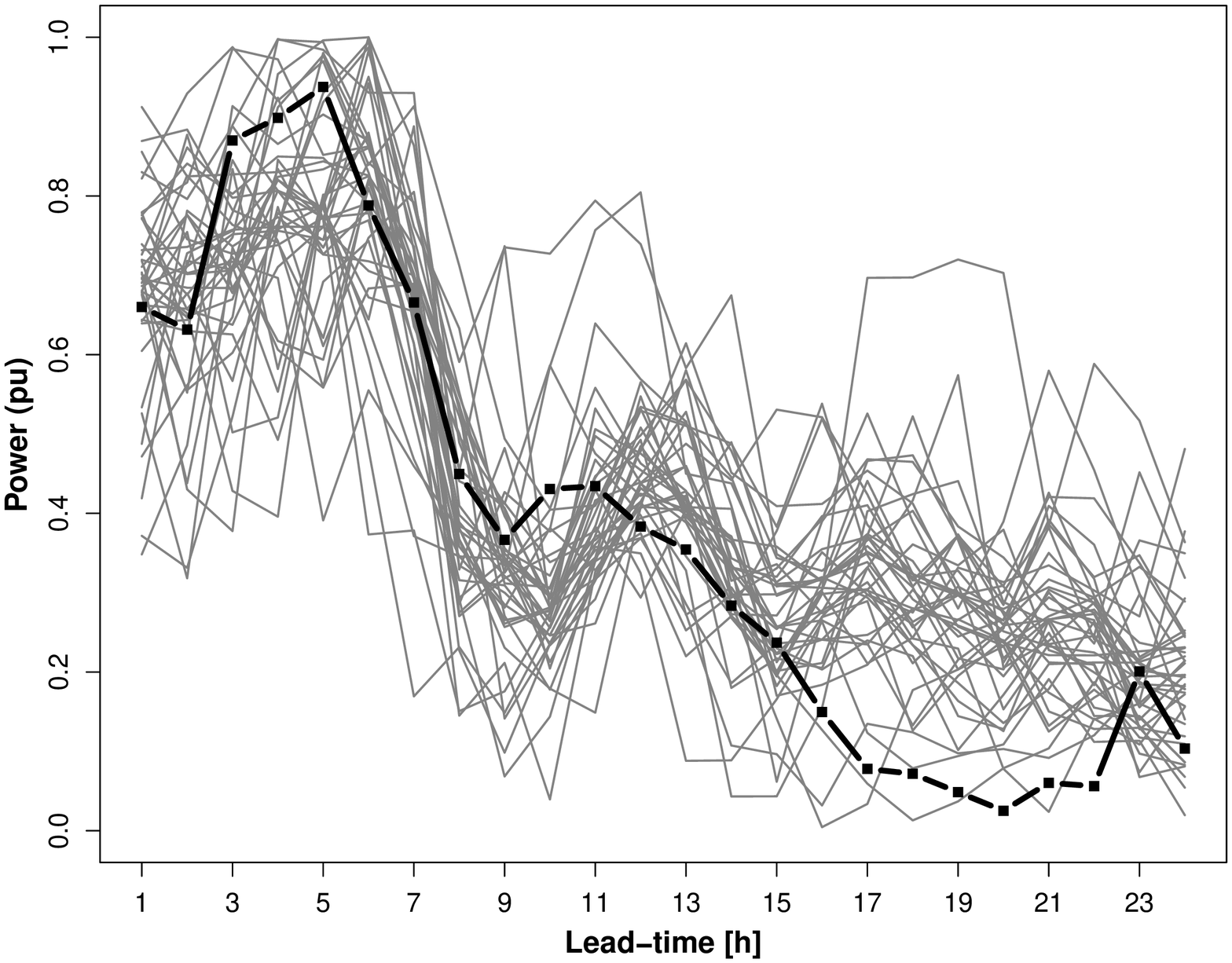}
			\caption{}
			\label{fig:Scenarios1}
		\end{subfigure}
		\begin{subfigure}{.33\textwidth}
			\centering
			\includegraphics[width=.9\linewidth]{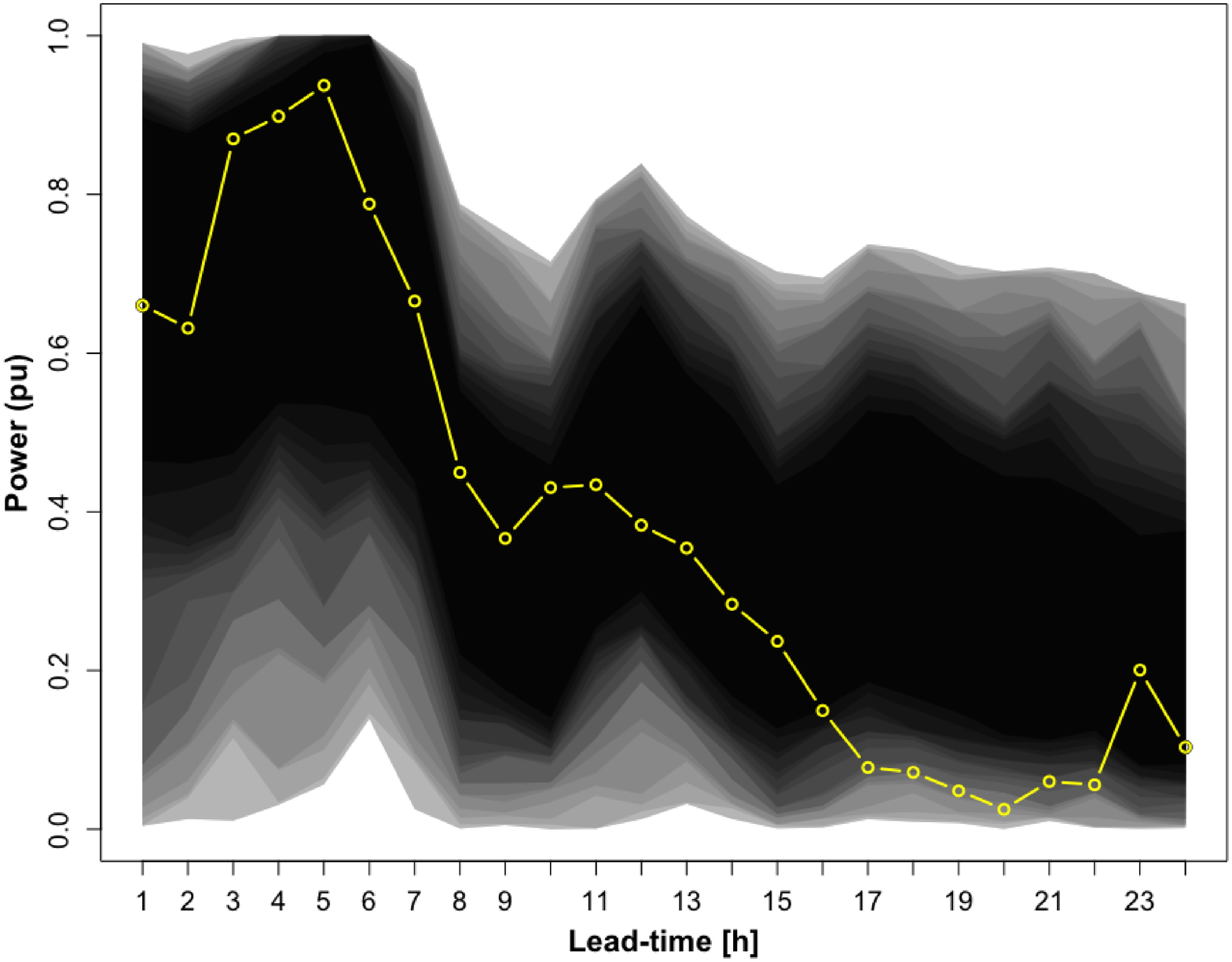}
			\caption{}
			\label{fig:MPIIntervals1}
		\end{subfigure}
	\end{tabular}
	\caption{(a) PV observations (yellow color curves) along with 19 univariate prediction intervals with coverage ranging from 0.05 to 0.95 in 0.05 increments (from the darkest to the lightest, (b) PV observations (dark black color curves) along with 40 generated space-time trajectories (gray color curves), (c) PV power observations (yellow color curves)  along with 19 MPIs with nominal coverage ranging from 0.05 to 0.95 in 0.05 increments (from the darkest to the lightest)}
	\label{fig:UPI_SC_MPI}
	\vspace{-0.5em}
\end{figure*}

For the bounded random variables, the size of the prediction polyhedra is determined
by the intersection of two polyhedra. The first one is the prediction polyhedron itself and the second one is formed by the feasible range of the random variable. For the case of  wind/PV power, the second polyhedron is a hyper-cube with edges of length equal to the maximum capacity of generation unit. Because there is no simple formulation  to calculate the intersection analytically,  we use a Monte Carlo-based method for estimation of the volume of prediction regions~\cite{golestaneh2018ellipsoidal}. The idea  is to generate $ N' $ random samples in the feasible range and then calculate the proportion of those points which lie in the prediction polyhedron. The volume of prediction polyhedra $V^P$ is calculated as
\begin{equation}
\label{VolumeGeneral}
V^P={N'' V^c}/{N'}
\end{equation}
with $ N'' $ as the number of $ D $-dimensional points enveloped by the prediction polyhedron and $ V^c $ is the volume of the bounded hyper-cube.
\begin{figure*}[h!]
	\begin{tabular}[c]{cc}
		\begin{subfigure}{.45\textwidth}
			\centering
			\includegraphics[width=\linewidth,height=5.0cm]{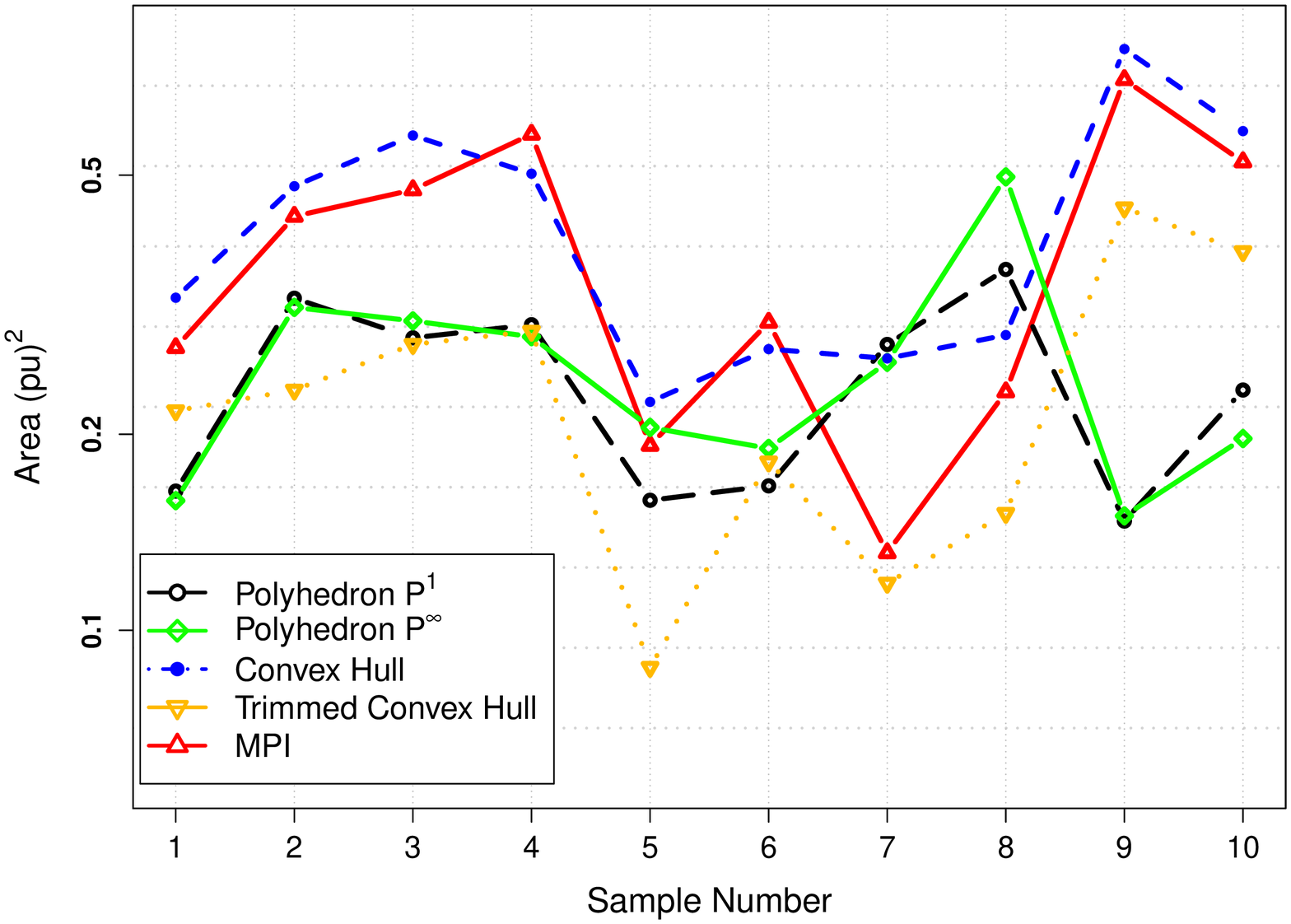}
			\caption{Dimension 2}
			\label{fig:Vol Dim2}
		\end{subfigure}
		
		\begin{subfigure}{.45\textwidth}
			\centering
			\includegraphics[width=\linewidth,height=5.0cm]{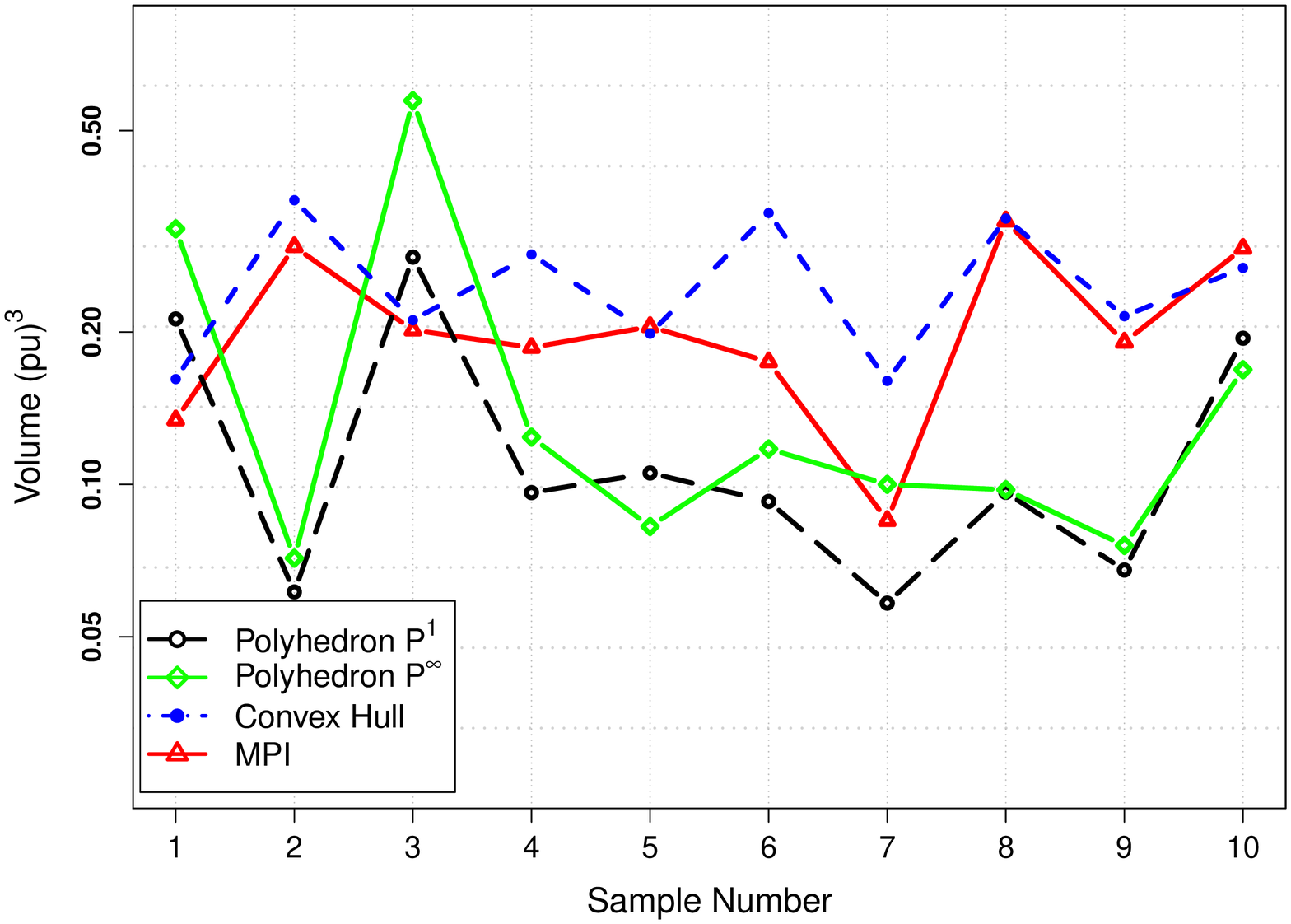}
			\caption{Dimension 3}
			\label{fig:Vol Dim 3}
		\end{subfigure}
		\\
		\begin{subfigure}{.45\textwidth}
			\centering
			\includegraphics[width=\linewidth,height=5.0cm]{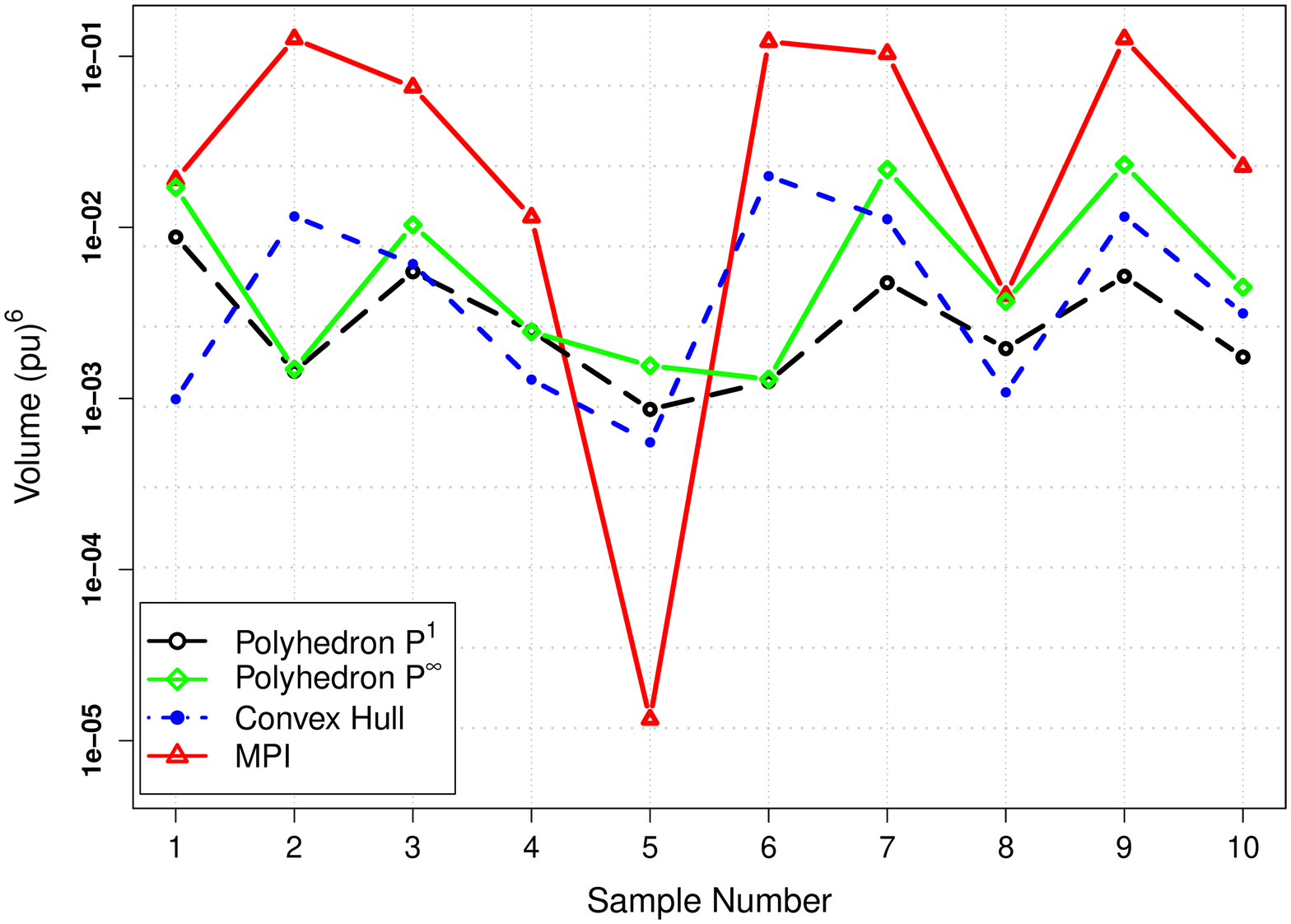}
			\caption{Dimension 6}
			\label{fig:Vol Dim 6}
		\end{subfigure}
		
		\begin{subfigure}{.45\textwidth}
			\centering
			\includegraphics[width=\linewidth,height=5.0cm]{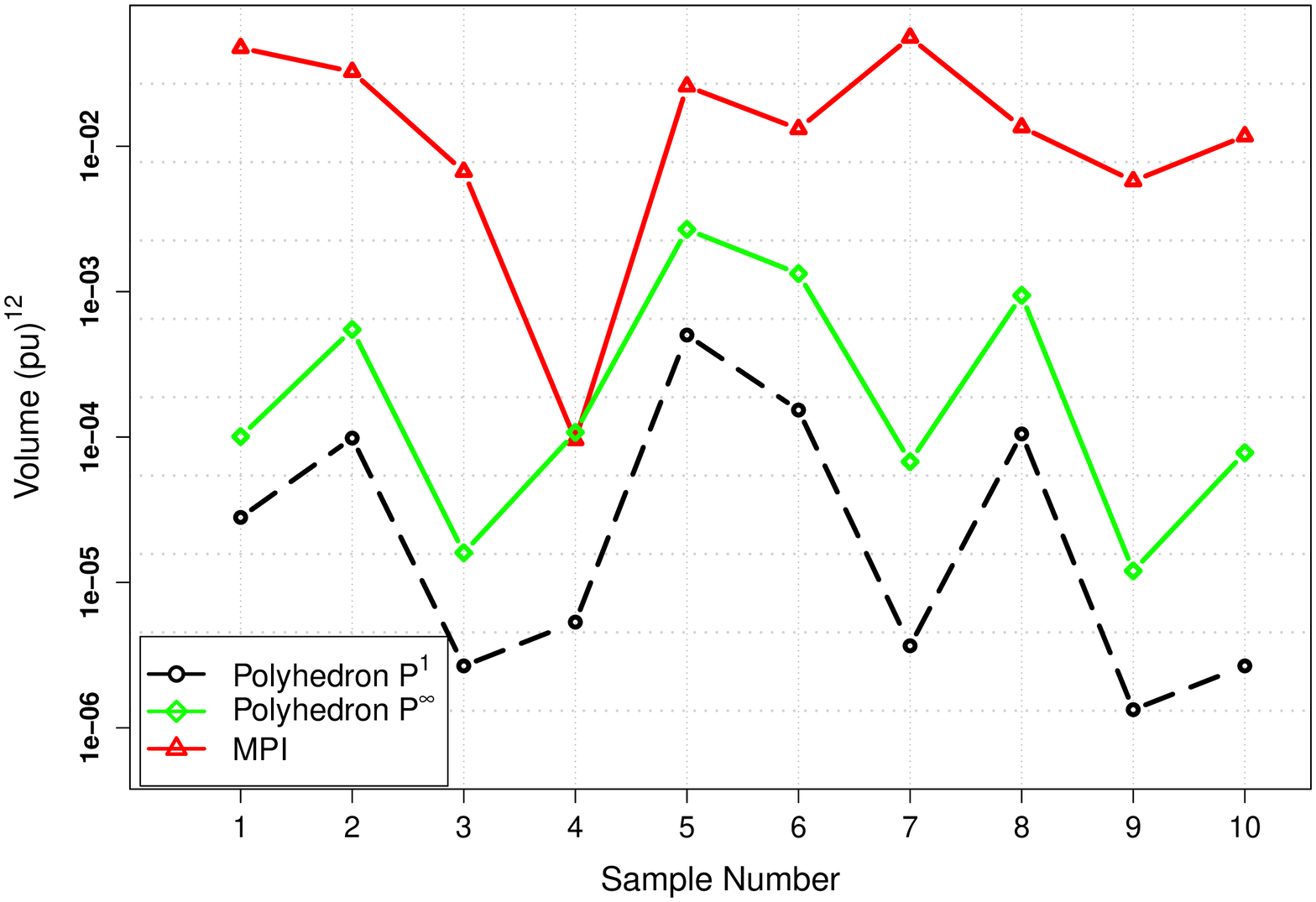}
			\caption{Dimension 12}
			\label{fig:Vol Dim 12}
			\vspace{-0.6em}
		\end{subfigure}%
	\end{tabular}
	\caption{Estimated volume of prediction polyhedra with nominal coverage rates (a) 95\%, (b) 90\%,  (c) 85\% and (d) 80\%. \label{fig:Volume}}
	\vspace{-0.5em}
\end{figure*}
Fig. \ref{fig:Volume} illustrates the size of the  proposed prediction polyhedra in comparison with MPI for ten randomly selected days from the evaluation data. The prediction polyhedra of sizes 2, 3, 6 and 12 and nominal probabilities 95\%, 90\%, 85\% and 80\% are included in the figure. It is to be noted that to study the results for different days, the   selected days are not the same for all  dimensions shown in the figure. The vertical axes  are logarithmic for  a clearer illustration. The empirical coverage of original prediction convex hulls is 90\% and it reduces to 87\% for trimmed convex hulls. However, as trimming the convex hulls in dimension 2 reduces their sizes significantly, we recommend discarding outliers regardless. Among the four techniques, $P^1$ shows the overall best performance in terms of conservativeness and sharpness. As the dimension increases, the MPIs become wider and more conservative. For example, in dimension 12 for the first day, 80\% MPI is more than 100 times larger than 80\% $P^1$. In order to better visualize the size of MPIs in higher dimensions, Fig. \ref{fig:UPI_SC_MPI} provides the MPIs for a randomly selected day from the  evaluation data along with the marginal prediction intervals and multivariate trajectories used as their inputs. As shown in the figure, MPIs specially those with low coverage rates are  wide and low in sharpness.

\begin{figure}[!t]
	\vspace{-0.4em}
	\centering
	\includegraphics[width=7cm,height=6.5cm]{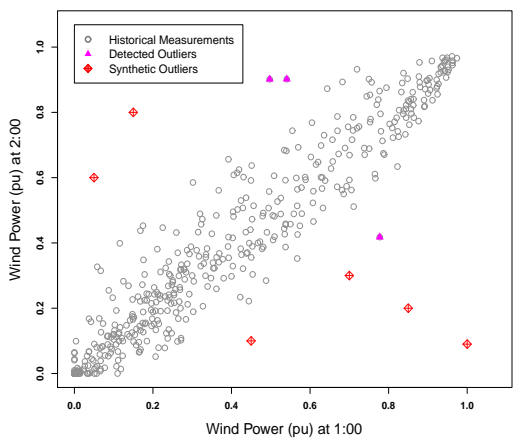}
	\caption{{Temporal scenarios predicted for a randomly selected day from the wind power data. Few scenarios are detected as outliers. 6 Synthetic outliers are generated for analysis.}}
	\label{Fig:SyntheticOutliers}
	\vspace{-0.4em}
\end{figure}
In almost all real world datasets, there is a possibility of outlier occurrence. Outliers might come from error in measurement, collection or communication of data. Outliers  can pose serious problems in statistical analysis and grossly distort and mislead them. The first measure to deal with outliers is to identify them, then either they can be discarded  or replaced  with more consistent data. Detection of multivariate outliers is discussed in subsection \ref{Subsection: Outliers}. It is important to assess the robustness of various regression models to outliers. Fig. \ref{Fig:SyntheticOutliers} shows historical wind power measurements available in the training subset recorded at 1:00 am and 2:00 am. Following the approach discussed in subsection \ref{Subsection: Outliers},  those observations with Mahalanobis distance higher than $1.1\chi^2_2 (\alpha=0.001)$ are detected as outliers. We generate six more synthetic outliers as shown in Fig. \ref{Fig:SyntheticOutliers} and substitute them for 6 randomly selected measurements. Then, we compare the performance of prediction polyhedra with and without the presence of those synthetic outliers. Based on empirical results, outliers on average change the empirical coverage of MPIs, $P^1$, $P^\infty$ and convex hulls by 1.5\%, 0.7\%, 0.9\% and 3\%, respectively. The occurrence of outlines also changes the volumes of 90\% prediction polyhedra by  0.048, 0.012, 0.020 and 0.044 for MPIs, $P^1$, $P^\infty$ and convex hulls, respectively. The results indicate that $P^1$ provides the highest robustness to outliers followed by $P^\infty$.  

\vspace{-0.6em}
\section{Conclusion}
\label{Section:Conclusion}	
	In order to facilitate the transition from deterministic forecasts to the point where end-users can confidently harness uncertainty information, it is required  to develop  frameworks allowing to characterize uncertainty in  forms that suit best the needs of various decision-making communities. Due to growing interests in polyhedral uncertainty sets,  this work  proposed frameworks to generate, calibrate and evaluate uncertainty information in the form of multivariate polyhedra for PV and wind power and within  various temporal and spatial scales. Two of the proposed techniques  use point and correlation matrix forecasts as inputs and predict the uncertainty budget such that prediction polyhedra provide  the desired probability levels and  conservativeness. Two other techniques  work based on finding convex hulls of spatial/temporal scenarios. The proposed approaches together with multivariate prediction intervals as a benchmark  are compared based on their calibration and conservativeness. The empirical results suggest that prediction convex hulls are not recommended for wind/PV predictions in dimensions higher than four. $P^1$ shows overall the best performance followed by $P^{\infty}$.  Both $P^1$ and $P^{\infty}$ are promising formulations for skilled  uncertainty characterization in convex forms and their performance does not degrade as the dimension increases, provided that their parameters are predicted appropriately. 




\vspace{-0.3em}
\bibliographystyle{IEEEtran}
\vspace{-0.34em}
\bibliography{reftest}
\end{document}